\documentclass[12pt, prd, showpacs]{revtex4}
\usepackage{amssymb}
\usepackage{amsmath}

\input{tcilatex}

\begin{document}

\title{On white holes as particle accelerators}
\author{O. B. Zaslavskii}
\affiliation{Department of Physics and Technology, Kharkov V.N. Karazin National
University, 4 Svoboda Square, Kharkov 61022, Ukraine}
\affiliation{Institute of Mathematics and Mechanics, Kazan Federal University, 18
Kremlyovskaya St., Kazan 420008, Russia}
\email{zaslav@ukr.net }

\begin{abstract}
We analyze scenarios of particle collisions in the metric of a nonextremal
black hole that can potentially lead to ultrahigh energy $E_{c.m.}$ in their
centre of mass frame. Particle 1 comes from infinity to the black hole
horizon while particle 2 emerges from a white hole region. It is shown that
unbounded $E_{c.m.}$ $\ $require that particle 2 pass close to the
bifurcation point. The analogy with collisions inside the horizon is
discussed.
\end{abstract}

\maketitle

\section{Introduction}

Several years ago, an interesting effect was discovered. It turned out that
if two particles collide near a rotating extremal black hole, the energy $%
E_{c.m.}$ in their centre of mass frame can become unbounded \cite{ban}.
This is called the Ba\~{n}ados-Silk-West (BSW) effect, after the names of
its authors. Later on, this effect was generalized to nonextremal horizons 
\cite{gp}, generic rotating black holes \cite{prd} and even nonrtotaing
charged ones \cite{jl}. In all these cases it is implied that both particles
move towards the horizon as usual for black holes.

Meanwhile, there are also scenarios with head-on collisions when one of
particles moves away from the horizon. They were mentioned cursorily in Sec.
II G of \cite{pir3} for the rotating case although the term "white hole" was
not used there. The detailed coherent treatment of this type of scenario was
done in \cite{w} where the role of white holes was stressed and it was
noticed that unbounded $E_{c.m.}$ appear even for the Schwarzschild metric.
As is known, the spacetime of an eternal black hole includes inevitably two
regions - black hole and white hole ones. In the scenario considered in \cite%
{w} particle 1 moves towards the future horizon whereas particle 2
approaches the past horizon from the inner white hole region. For
nonextremal horizons (which is the subject of the present work) this means
in terms of R- and T-regions \cite{rt} that particle 2 passes from the
expanding T-region to the R one whereas particle 1 moves within our R region
as usual. This scenario works for generic particles in contrast to the
standard BSW effect where fine tuning between parameters of one particle is
required and is valid for generic eternal black - white holes.

In the present work, we analyze this scenario further, describe the main
features of relevant trajectories and argue that there exists close
similarity between such a scenario and high energy collision inside the
horizon.

Some reservations are in order. The existence of white wholes in
questionable. In particular, they can be unstable (see Sec. 15 of \cite{fn}%
). However, we can point at least to three factors that support our
motivation. (i) Many years ago, an interesting conjecture was pushed forward
according to which white holes can act as region retarded in the expansion
of surrounded matter in Universe \cite{del}. It is important that the
scenario considered there includes, in particular, collision between
particles that leave a white hole and those that move outside that
corresponds just to our case. (ii) Typically, the structure of spacetime
includes alternation of R and T regions. For example, this happens for
regular black holes, so-called black universes \cite{br}, the motion of
self-gravitating shells \cite{bd}, etc. (iii) Even if (i) and (ii) are not
realized in astrophysics in practice, collisions of particles near white
holes is an essential ingredient of the theory of high energy collisions.
Without this treatment, our understanding of the BSW effect and its
modifications would remain incomplete. It is also worth noting that the
energetics of white holes was discussed \ along time ago but in a quite
different context \cite{nar}.

Throughout the paper, we use systems of units in which fundamental constants 
$G=c=1.$

\section{Basic equations}

Let us consider the metric of the eternal hole%
\begin{equation}
ds^{2}=-dt^{2}f+\frac{dr^{2}}{f}+r^{2}(d\theta ^{2}+\sin ^{2}\theta d\phi
^{2})\text{,}  \label{m}
\end{equation}%
where $f(r_{+})=0$. For the Schwarzschild metric $f=1-\frac{r_{+}}{r}$. We
consider pure radial motion. Then, for a free particle having the mass $m$
equations of motion read

\begin{equation}
\dot{r}=\sigma \sqrt{\varepsilon ^{2}-f}\text{,}  \label{dr}
\end{equation}%
\begin{equation}
\frac{dt}{d\tau }=\frac{\varepsilon }{f}\text{,}
\end{equation}%
\begin{equation}
\frac{dr}{dt}=\sigma \frac{f\sqrt{\varepsilon ^{2}-f}}{\varepsilon }.
\label{rt}
\end{equation}%
Here, $\varepsilon =\frac{E}{m},$ $E$ is the energy, $\sigma =\pm 1$
depending on the direction of motion, dot denotes differentiation with
respect to the proper time $\tau $.

Let two particles collide. One can define the energy in the centre of mass
in the point of collision according to%
\begin{equation}
E_{c.m.}^{2}=-P_{\mu }P^{\mu }\text{,}
\end{equation}%
$P^{\mu }=m_{1}u_{1}^{\mu }+m_{2}u_{2}^{\mu }$, where $u^{\mu }$ is the
four-velocity. Then,%
\begin{equation}
E_{c.m.}^{2}=m_{1}^{2}+m_{2}^{2}+2m_{1}m_{2}\gamma \text{,}
\end{equation}%
$\gamma =-u_{1\mu }u_{2}^{\mu }$ is the Lorentz factor of relative motion.

Let particle 1 with $\sigma =-1$ and particle 2 with $\sigma =+1$ collide at 
$r=r_{c}$. (Hereafter, subscripts "c" implies that $r=r_{c}$.) Then, it
follows from the equations of motion that%
\begin{equation}
\gamma =\frac{\varepsilon _{1}\varepsilon _{2}+\sqrt{\varepsilon
_{1}^{2}-f_{c}}\sqrt{\varepsilon _{2}^{2}-f_{c}}}{f_{c}}\text{,}  \label{ga}
\end{equation}%
where $f_{c}=f(r_{c})$. If collision happens close to the horizon, so $%
r_{c}\rightarrow $ $r_{+}$, the quantity $f_{c}\rightarrow 0$ and we obtain
formally diverging expression.

However, there is an essential subtlety here, not discussed in \cite{pir3}, 
\cite{w}. The effect under consideration involves not one horizon as usual
in the BSW effect but two different horizons - the future (black hole)
horizon and the past (white hole) one. In such a situation a new problem
arises that remained irrelevant for collisions near a black hole horizon
only. To gain the effect, one should guarantee, first of all, that collision
does occur near the horizon. Otherwise, either particle approaches its own
horizon in different points of the spacetime diagram (see Fig. 1) and no
near-horizon collision happens. It is possible somewhere far from the
horizon but this case is uninteresting since the gamma factor $\gamma $
remains modest.\FRAME{ftbpFU}{2.6429in}{2.1127in}{0pt}{\Qcb{Generic
trajectories do not intersect near the horizon.}}{}{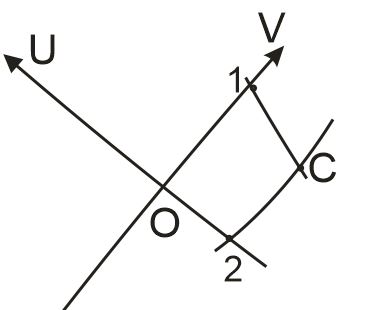}{\special%
{language "Scientific Word";type "GRAPHIC";display "USEDEF";valid_file
"F";width 2.6429in;height 2.1127in;depth 0pt;original-width
1.2566in;original-height 0.8934in;cropleft "0";croptop "1";cropright
"1";cropbottom "0";filename 'fig1.JPG';file-properties "XNPEU";}}

To elucidate the essence of matter, it makes sense to introduce the Kruskal
coordinates that cover all spacetime including the black and white hole
regions. Then, according to the standard formulas, we use in the region $%
r>r_{+}$ (our R region) coordinates 
\begin{equation}
U=-\exp (-\kappa u)\text{, }V=\exp (\kappa v),  \label{UV}
\end{equation}%
\begin{equation}
u=t-r^{\ast }\text{, }v=t+r^{\ast }\text{,}  \label{uvr}
\end{equation}%
where the so-called tortoise coordinate equals%
\begin{equation}
r^{\ast }=\int^{r}\frac{dr}{f}\text{,}  \label{tor}
\end{equation}%
$\kappa $ is the surface gravity. From (\ref{UV}) - (\ref{tor}) useful
relation follow:%
\begin{equation}
UV=\exp (2\kappa r^{\ast })\text{,}  \label{prod}
\end{equation}%
\begin{equation}
\frac{V}{\left\vert U\right\vert }=\exp (2\kappa t)\text{.}  \label{rat}
\end{equation}

Near the horizon,%
\begin{equation}
r^{\ast }\approx \frac{1}{2\kappa }\ln \frac{r-r_{+}}{r_{+}}+C\text{,}
\end{equation}%
where $C$ is a constant,%
\begin{equation}
f\approx 2\kappa r_{+}UV\text{.}
\end{equation}

For the Schwarzschild metric, with the constant of integration chosen
properly, one finds exact expressions%
\begin{equation}
r^{\ast }=r+r_{+}\ln \frac{r-r_{+}}{r_{+}}\text{,}
\end{equation}%
\begin{equation}
\kappa =\frac{1}{2r_{+}},  \label{k}
\end{equation}

\begin{equation}
f=\frac{r_{+}}{r}\exp (-\frac{r}{r_{+}})UV.
\end{equation}

Then, the metric takes the form%
\begin{equation}
ds^{2}=-FdUdV+r^{2}(d\theta ^{2}+\sin ^{2}\theta d\phi ^{2})\text{,}
\end{equation}%
\begin{equation}
F=f\frac{du}{dU}\frac{dv}{dV}=-\frac{f}{UV\kappa ^{2}}\text{.}
\end{equation}

We must have in the point of collision%
\begin{equation}
U_{1}=U_{2}\text{, }V_{1}=V_{2}\text{.}  \label{uv}
\end{equation}

In terms of the $t$ coordinate,%
\begin{equation}
t_{1}(r_{c})=t_{2}(r_{c})\text{.}  \label{t12}
\end{equation}%
Here, according to (\ref{rt}),%
\begin{equation}
t_{1}(r)=\varepsilon _{1}\int_{r}^{r_{1}}\frac{dr}{f\sqrt{\varepsilon
_{1}^{2}-f}}\text{,}  \label{t1}
\end{equation}%
where $r_{1}$ is the starting point of motion of particle 1, so $%
t_{1}(r_{1})=0$. In a similar way,%
\begin{equation}
t_{2}(r)=t_{1}(r_{c})-\varepsilon _{2}\int_{r}^{rc}\frac{dr}{f\sqrt{%
\varepsilon _{2}^{2}-f}}\text{.}  \label{t2}
\end{equation}

When particle 2 crossed the past horizon, $r\rightarrow r_{+}$ and $%
t_{2}\rightarrow -\infty $ that signals about failure of the original
coordinate system (\ref{m}). However, the proper time $\tau $ stays finite,
so particles 1 and 2 can meet in the point $r_{c}$.

The choice of the constant of integration in (\ref{t2}) ensures that (\ref%
{uv}) is satisfied. If particle 1 comes from infinity, $\varepsilon \geq 1$.
It terms of the Kruskal coordinates, the equations of motion follow from (%
\ref{dr}) - (\ref{rt}) and (\ref{UV}) - (\ref{tor}). They read%
\begin{equation}
\frac{dU}{dr}=-\frac{\sigma \kappa U}{\sqrt{\varepsilon ^{2}-f}(\varepsilon
+\sigma \sqrt{\varepsilon ^{2}-f})},  \label{Ur}
\end{equation}%
\begin{equation}
\frac{dV}{dr}=\frac{\kappa V(\sqrt{\varepsilon ^{2}-f}+\sigma \varepsilon )}{%
f\sqrt{\varepsilon ^{2}-f}}\text{.}  \label{Vr}
\end{equation}

\section{Kinematics of collision}

In the $(U,V)$ coordinate system, particle 2 crosses the white hole horizon
in the point $(U_{2},0)$. Particle 1 would cross the black hole horizon in
the point $(0,V_{1})$, unless the collision happened. In general, their
trajectories intersect in the intermediate point with $\left\vert
U_{c}\right\vert =O(1)$, $V_{c}=O(1)$ this gives a modest $\gamma .$ To gain
large $E_{c.m.}$, we must arrange collision very nearly to the horizon,
where $f_{c}$ is small, so $\gamma $ is big according to (\ref{ga}). As in
the present work we are interested in the effects near the white hole
horizon $V=0$, we require 
\begin{equation}
V_{c}\ll 1\text{.}  \label{vsm}
\end{equation}

This entails consequences for the properties of a trajectory of each
particle.

\subsection{Particle 1}

By assumption, particle 1 started its motion at $t_{1}=0$. We can choose,
say, that for $t<0$ it remained in the state of the rest, $r=r_{1}=const$.
Then, $t>0$ on its further trajectory. It is seen from (\ref{rat}) that $%
\left\vert U_{c}\right\vert <V_{c}$. This means that collision could not
happen near a generic point of the white horizon where $U=O(1)$, $V=0$ since
this would have been in contradiction with (\ref{rat}) and (\ref{vsm}). As
now both $\left\vert U_{c}\right\vert \ll 1$ and $V_{c}\ll 1$, the collision
happens near the bifurcation point $V=0=U$.

\subsection{Particle 2}

Let us consider particle 2 moving from a white hole. It has $\sigma =+1$. We
want to arrange collision near the past horizon, so $r_{c}-r_{+}$ is small.
Then, we have%
\begin{equation}
U_{c}\approx U_{+}+\left( \frac{dU}{dr}\right) _{+}(r_{c}-r_{+})\approx
U_{+}-\frac{\kappa U_{+}}{2\varepsilon _{2}^{2}}(r_{c}-r_{+})\approx U_{+}(1-%
\frac{f_{c}}{4\varepsilon _{2}^{2}})\text{,}  \label{uf}
\end{equation}%
where $U_{+}=U(r_{+})$ and we used the fact that near the horizon%
\begin{equation}
f(r)\approx 2\kappa (r-r_{+})\text{.}  \label{fk}
\end{equation}%
For any finite $\varepsilon _{2}$, this gives a small correction to $U_{+}$,
so $U_{c}\approx U_{+}$. Thus both the point where particle 2 intersects the
horizon and the point of collision are situated near the bifurcation point.

For completeness, we will also discuss the case when both $\varepsilon
_{2}^{2}$ and $f_{c}$ are small and have the same order, 
\begin{equation}
\varepsilon _{2}^{2}\sim f_{c}\text{,}
\end{equation}%
so the Taylor expansion (\ref{uf}) does not work. In the near-horizon region
eq. (\ref{Ur}) for particle 2 gives us%
\begin{equation}
\frac{d\ln \left\vert U\right\vert }{dr}\approx -\frac{\kappa }{\sqrt{%
\varepsilon _{2}^{2}-2\kappa (r-r_{+})}}\frac{1}{\sqrt{\varepsilon
_{2}^{2}-2\kappa (r-r_{+})}+\varepsilon _{2}}\text{,}  \label{dur}
\end{equation}%
where we took into account (\ref{fk}). It is convenient to use
parametrization $\varepsilon _{2}^{2}=2\kappa (r_{0}-r_{c})$. Here, $r_{0}$
is close to $r_{c}$ which, in turn, is close to $r_{+}$. Collision should
occur before particle 2 reaches the turning point, otherwise $\sigma $ would
change the sign and the head-on collision would not occur. Therefore, $%
r_{c}\leq \frac{r_{+}+r_{0}}{2}$.

Then, after integration with the boundary condition $U(r_{c})=U_{c}$, one
finds that%
\begin{equation}
U\approx \frac{U_{c}(\sqrt{\delta }+\sqrt{\delta -x})}{\sqrt{\delta }+\sqrt{%
\delta -\alpha }}\text{,}
\end{equation}%
where $\delta =\frac{r_{0}-r_{c}}{r_{+}}$, $x=\frac{r-r_{+}}{r_{+}}$ and $%
\alpha =$ $\frac{r_{c}-r_{+}}{r_{+}}=x(r_{c})$. To have radicals
nonnegative, we require $\alpha \leq \delta $. Thus%
\begin{equation}
U_{+}\approx U_{c}\frac{2}{1+\sqrt{1-\frac{\alpha }{\delta }}}\text{.}
\end{equation}

We see that $U_{+}$ and $U_{c}$ have the same order, so both of them are
small according to the explanation given above. Thus in both cases $%
\varepsilon _{2}\gg f_{c}$ and $\varepsilon _{2}\sim f_{c}\ll 1$ collision
occurs near the bifurcation point. See Fig. 2. where the scale is magnified
to show the whole picture distinctly.

\FRAME{ftbpFU}{2.4941in}{2.636in}{0pt}{\Qcb{Collision near the bifurcation
point.}}{}{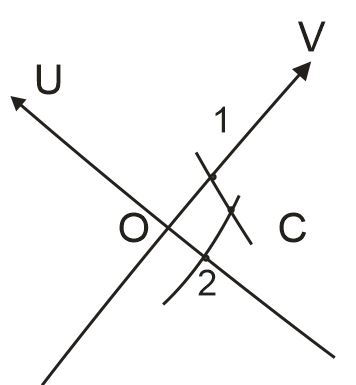}{\special{language "Scientific Word";type
"GRAPHIC";display "USEDEF";valid_file "F";width 2.4941in;height
2.636in;depth 0pt;original-width 1.2566in;original-height 0.8934in;cropleft
"0";croptop "1";cropright "1";cropbottom "0";filename
'fig2.JPG';file-properties "XNPEU";}}

It is worth noting that if $\varepsilon _{2}\sim \sqrt{f_{c}}$, $\varepsilon
_{1}=O(1)$, $\gamma =O(f_{c}^{-1/2})$, so the growth of $E_{c.m.}$ is more
slow than in the case $\varepsilon _{2}\gg \sqrt{f_{c}}$ when $\gamma
=O(f_{c}^{-1})$. If both particles have energies $\varepsilon _{1}\sim
\varepsilon _{2}\sim \sqrt{f_{c}},\gamma =O(1)$ and the effect of high
energy collision is absent.

Formally, there exists one more case when particle 2 falls back into a black
hole. However, collisions near the black hole horizon were already discussed
intensively in literature and it is known that without a fine-tuned
(critical) particle high $E_{c.m.}$ are impossible \cite{ban} - \cite{prd}.
Therefore, we do not consider this case.

\section{Discussion and conclusions}

If two particle moving along the line in opposite directions collide in the
flat spacetime, one can arrange collision in any given point adjusting an
initial position of, say, particle 2 to its energy. However, we saw that for
head-on collision near the white-black hole horizons the situation is
different. The particle emerging from a white hole region should pass close
to the bifurcation point, although not through the bifurcation point itself
since otherwise it would come into the contracting T region instead of our R
one. And, this is true irrespective of the energy of particle 2. Collision
of both particles 1 and 2 also happens near the bifurcation point.

It is instructive to compare the results with those found for collisions
inside black (white) holes since in both cases one is faced with the
existence of two branches of the horizon. At first, it was claimed in \cite%
{lake1} that collisions near the inner nonextremal horizon lead to the
unbounded growth of $E_{c.m.}$ in a manner similar to collisions near the
event horizon of rotating black holes \cite{ban}. Later on, this result was
refuted \cite{lake2} because of impossibility to arrange collision
kinematically since each particle approaches to its horizon. The similar
conclusion was made in the end of Sec. 3 in \cite{gp-astro}. However, more
careful treatment showed that the effect of high $E_{c.m.}$ can be saved 
\cite{inner} - \cite{flat} if particles pass very close to the bifurcation
point. We see that similar situation happens in \ the present case although
the number of suitable scenarios actually reduced to one. This is because
kinematics of the problem is more restricted (particle 1 moves in the R
region only, particle 2 moves from a white hole to the R region).

Thus white holes in combination with the black ones can serve as
accelerators of particles to ultra-high energies. In contrast to the
standard BSW effect \cite{ban}, no fine-tuning of parameters is required
but, instead, there is a kinematic restriction. This is necessary if we want
to achieve unbounded $E_{c.m.}$

\begin{acknowledgments}
This work was funded by the subsidy allocated to Kazan Federal University
for the state assignment in the sphere of scientific activities. I
acknowledge also support from SFFR,
\end{acknowledgments}

\begin{acknowledgments}
Ukraine, Project No. 32367
\end{acknowledgments}

\end{document}